\documentstyle[12pt,aaspp4,amstex]{article}

\begin{document}

\title{The Epoch of Helium Reionization}

\author{Aaron Sokasian\footnote{asokasia@@cfa.harvard.edu}}
\author{Tom Abel\footnote{hi@@tomabel.com} and 
Lars Hernquist\footnote{lars@@cfa.harvard.edu}} 
\affil{Department of Astronomy, Harvard University,
Cambridge, MA 02138}
\authoremail{asokasia@cfa.harvard.edu}

\begin{abstract}

We study the reionization of Helium {\small II} by quasars using a
numerical approach that combines 3D radiative transfer calculations
with cosmological hydrodynamical simulations.  Sources producing the
ionizing radiation are selected according to an empirical quasar
luminosity function and are assigned luminosities according to their
intrinsic masses.  The free parameters associated with this procedure
are: (1) a universal source lifetime, (2) a minimum mass cutoff, (3) a
minimum luminosity cutoff, (4) a solid angle specifying the extent to
which radiation is beamed, and (5) a tail-end spectral index for the
radiative energy distribution of the sources.  We present models in
which these parameters are varied and examine characteristics of the
resultant reionization process that distinguish the various cases.   In
addition, we extract artificial spectra from the simulations and
quantify statistical properties of the spectral features in each
model.

We find that the most important factor affecting the evolution of He
{\small II} reionization is the cumulative number of ionizing photons
that are produced by the sources.  Comparisons between He {\small} II
opacities measured observationally and those obtained by our analysis
reveal that the available ranges in plausible values for the
parameters provide enough leeway to provide a satisfactory match.
However, one property common to all our calculations is that the epoch
of Helium {\small II} reionization must have occurred at a redshift
between $3\lesssim z \lesssim 4$.  If so, future observational
programs will be able to directly trace the details of the ionization
history of helium and probe the low density phase of the intergalactic
medium during this phase of the evolution of the Universe.

\end{abstract}


\section{INTRODUCTION}

Space-based ultraviolet telescopes have made it possible to observe
the Ly$\alpha$ transition of singly ionized helium (He {\small II})
along lines of sight to high-redshift quasars. The discovery of the
``He {\small II} Gunn-Peterson effect'' in Q0302-003 ($z=3.285$) by
Jakobsen et al. (1994) marked the beginning of He {\small II} Ly$\alpha$
studies. He {\small II} absorption in a second quasar, PKS 1935-692
($z=3.18$), was identified by Tytler (1995; see also Jakobsen 1996),
while Davidsen et al. (1996) measured the He {\small II} opacity at a
lower redshift of $z=2.72$ towards the quasar HS 1700+6416 using the
Hopkins Ultraviolet Telescope (HUT). These papers established the
presence of He {\small II} absorption, albeit at
low resolution, and indicated that the
mean opacity increases with redshift. Later, Q0302-003 and HE 2347-4342
($z=2.885$) were observed individually by Hogan, Anderson \& Rugers
(1997) and Reimers et al. (1997) using the Goddard High Resolution
Spectrograph (GHRS). Subsequently, Heap et al. (2000) and Smette et
al. (2000) reported new HST/STIS spectra for Q0302-003 and HE
2347-4342, respectively. These observations had sufficient resolution to
begin to resolve some of the features in the He {\small II} Ly$\alpha$
forest and also allowed them to to cross-correlate the data
with the absorbers
in the gaps of the hydrogen (H {\small I}) forest lines. In
particular, the high quality of the STIS spectra reveals regions of
high-opacity as well as ones of low opacity extending over several
Mpc. Moreover, models of the spectra based on corresponding H {\small
I} Ly$\alpha$ forests were presented to probe the hardness of the UV
background, believed to emanate from the observed quasar population
(e.g. Haardt \& Madau 1996).

Due to its relatively higher optical depth compared to H {\small I},
He {\small II} serves as a better probe of diffuse gas residing in
voids between galaxies (e.g. Croft et al. 1997).  Resolving He {\small
II} absorption features in quasar spectra can therefore reveal the
presence of matter in large, low density regions, which, according to
gravitational instability models of structure formation, harbor the
bulk of the baryonic matter in the universe at high
redshift (e.g. Dav\'e et al. 2001; Croft et al. 2001). 
Interpreting observations of He {\small II} absorption will
require comparisons with detailed models derived from cosmological
hydrodynamical simulations which incorporate radiative transfer
effects responsible for the photoionization of helium.  In particular,
it is of interest to develop an understanding of how individual
sources act collectively in the reionization process.

In this paper we use the numerical approach described in Sokasian,
Abel \& Hernquist (2001, hereafter Paper I) to simulate the 3D
reionization of He {\small II} by quasars. In particular, we explore
the parameter space associated with the characteristics of the sources
and study how they influence global properties of the reionization
process.  Comparisons with observational results are also made
possible by extracting synthetic spectra from the simulations.  Here
our aim is twofold: to develop an understanding of the sensitivity of
the reionization process to source properties and to examine the
predictions of the different models in light of recent observational
results.

\section{SIMULATING COSMOLOGICAL REIONIZATION}

In general, a description of the evolution of ionization zones around
cosmological sources requires a full solution to the radiative
transfer equation.  Such a solution would yield everywhere the
monochromatic specific intensity of the radiation field in an
expanding universe: $I_{\nu}\equiv I(t,\vec{x},\hat{n},\nu)$, where
$\hat{n}$ is a unit vector along the direction of propagation of a ray
with frequency $\nu$.  Presently, it is computationally impractical to
acquire a complete, multi-dimensional solution for $I_{\nu}$ at the
high resolution required for cosmological simulations.  Our approach
relies on a new algorithm which employs a simple jump condition to
compute all radiative ionizations from a given source in a single
step.  This eliminates the need to repeatedly re-cast rays and
calculate rates at every time step, thereby greatly speeding up the
process.  Moreover, our algorithm includes an approximate treatment of
the diffuse component of the radiation field.

The approach we take is to approximately describe the evolution of the
ionization state of the gas in a cosmological volume by iteratively
calculating the net effect of ionizing sources at regularly spaced
time intervals via a ray casting scheme. At the end of each interval,
the effect of a diffuse ionizing background is accounted for by
casting rays of ionizing radiation (according to a detailed
prescription) inward from the sides of the simulation box. Density
fields as well as information regarding sources will be specified from
outputs at desired redshifts from a cosmological simulation. For
simplicity, all sources are taken to have identical lifetimes. The
details of the implementation of this method are discussed in Paper I.

Our approach requires a number of straightforward approximations to
simplify the calculations; we review them here.  First, our radiative
transfer calculations are done on a uniform Cartesian grid whose scale
$L$ will always be much smaller than the horizon, $c/H(t)$, where $c$
is the speed of light and $H(t)$ is the time dependent Hubble
constant.  This eliminates the need to include Doppler shifting of
frequencies in line transfer calculations.  Additionally, if the light
crossing time $L/c$ is much shorter than the ionization timescale, the
time dependence of the intensities drop out as well.  In the volumes
we simulate, this will certainly be true and so we make this
approximation as well.  Next, we assume that the density field will
experience negligible cosmological evolution during the lifetime of a
single source.  This requires us to consider only short-lived sources
(at most a few $\times 10^{7}$ years).  This assumption allows us to
perform all our radiative calculations during a source's lifetime in a
static density field, greatly reducing the complexity of the
algorithm.

Finally we ignore the dynamical consequences of thermal feedback into
the gas from radiative ionization, enabling us to decouple our
calculation of the radiation field from the hydrodynamical evolution
of the gas. This allows us to use existing outputs from cosmological
simulations to describe the evolving density field during the
reionization process. In reality, photoheating introduces extra heat
into the medium, and as ionization fronts (I-fronts) move from small
scales to large scales, there is a corresponding transfer of power
from small to large scales through nonlinear evolution. This effect is
somewhat accounted for in the underlying cosmological simulation used
in this paper which includes a uniform ionizing background capable of
heating the gas. However, one also expects additional heating due to
radiative transfer effects during the reionization process, and such
uniform backgrounds cannot reproduce the observed increase in gas
temperatures from the extra heating (Abel \& Haehnelt 1999 and
references therein). As a result, we expect minor systematic errors to
be present in our solutions. However, the main purpose of this paper
is to describe the general morphological evolution of the reionization
process on large scales, which is insensitive to these systematic
errors. In particular, recombination rates depend weakly on
temperature and hence a proper accounting of photons is possible even
if the temperatures are not computed precisely.

The cosmological simulation we will use in our analysis is based on a
smoothed particle hydrodynamics (SPH) treatment, computed using the
parallel tree-code GADGET developed by Springel, Yoshida \& White
(2001).  The particular cosmology we examine is a $\Lambda$CDM model
with $\Omega_{b}=0.04$, $\Omega_{DM}=0.26$, $\Omega_{\Lambda}=0.70$,
and $h=0.67$ (see, e.g., Springel, White \& Hernquist 2001).  The
simulation uses $224^{3}$ SPH particles and $224^{3}$ dark
matter particles in a $67.0 \ \text{Mpc}/h$ comoving box, resulting in
mass resolutions of $2.970\times 10^8 \ M_{\odot}/h$ and $1.970\times
10^9 \ M_{\odot}/h$ in the gas and dark matter components,
respectively.  The gas can cool radiatively to high overdensity (e.g.
Katz, Weinberg \& Hernquist 1996) and is photoionized by a diffuse
radiation field which is assumed to be of the form advocated by Haardt
\& Madau (1996; see also, Dav\'e et al. 1999).  When sources are
included in our treatment of helium reionization, the ionization state
of the helium is recalculated, ignoring the diffuse background that
was included in the hydrodynamical simulation (see Paper I for 
details).\footnote{We note that Springel \& Hernquist (2001) have 
recently pointed out that for the specific hydrodynamical treatment 
we employ the amount of gas in cold, dense clumps is subject to
some inaccuracy.  In our calculations, however, these uncertainties
are folded into our parameterization of the mass-to-light ratio
and do not affect our ionization analysis.}

\section{SOURCE SELECTION AND MODELS}

The full details of our source selection method are described in \S
4.2 of Paper I; here, we summarize the main steps and point out the
modifications adopted for this paper.  The basic procedure involves
identifying dense clumps of gas particles in the cosmological
simulation which represent plausible quasar locations, and adopting a
prescription for selecting a subset of these objects as actual sources
according to an empirical quasar luminosity function. The first task
required is to store in descending order an array of redshifts,
$z_{i}$, corresponding to intervals of one source lifetime,
$T_{life}$.  These redshifts will designate when sources turn on and
off during the ionization calculation. Next, we loop over the
cosmological simulation outputs, hereafter referred to as hydro
outputs, to identify plausible sources.  To identify quasars,
we specify a maximum linking length corresponding to a group
overdensity of roughly $200$. Additionally, we specify a minimum mass,
$M_{min}$, required for a group to be considered a source. For every
possible source identified, the center of mass location and total gas
mass of the group are recorded and stored in arrays.

Once a list of plausible sources for a given hydro output has been
compiled, all identified groups (sources) are logarithmically binned
by mass and a tally of the number of groups in each bin is made. The
results can then be fitted to form an analytic representation of the
mass function. Here we use a simple power-law of the form:
\begin{equation}
\frac{dN}{d\log{M}}=10^{(a\log{M}+b)}
\end{equation}   
where $dN$ is the number of groups with total gas masses between
$\log{M}$ and $\log{(M+dM)}$, and $a$,$b$ are fitted parameters.  Our
analytic representation is motivated by the general power-law nature
of theoretical halo mass functions (see Press \& Schechter 1974) and
seems to produce reliable fits for the range of group masses found in
our simulation volume. Due to the finite size of our volume, we
inevitably fail to capture the most massive groups ($>10\times
10^{12}\ M_{\odot}$) expected to exist on the exponential end of a
theoretical galaxy mass function. However, the number of sources that
would be selected off this end in our volume is less than unity on
average, therefore we do not expect that their exclusion will
significantly alter the results. Furthermore, the sources which are
selected with our evolving mass-to-light ratio roughly reproduce the 
observed range of quasar luminosities at the corresponding redshifts,
reassuring us that our cosmological volume is sufficiently large
enough to properly sample a representative subset of group masses.

Our next step is to define a selection criterion which will determine
a realistic subset of sources that will actually be activated during
the calculation of the radiation field.  In particular, we would like
to select sources according to the observed quasar luminosity function
$\phi(L,z)$ (LF). Here, $\phi(L,z)\ dL$ is the number of quasars
per unit comoving volume at redshift $z$, with intrinsic luminosities
between $L$ and $L+dL$.  Given the observational data, the inferred
luminosity function will depend upon the assumed cosmology and quasar
spectra. In our simulations, we will adopt the double-power-law model
presented by Boyle et al. (1988) using the open-universe fitting
formulae from Pei (1995) for the $B$-band (4400 \AA \ rest-wavelength)
LF of observed quasars, with a rescaling of luminosities and volume
elements for our $\Lambda$CDM cosmology. The parametric form of the
double power-law LF can be written as
\begin{equation}
\phi(L,z)=\frac{\Phi_{\star}/L_{z}}{(L/L_{z})^{\beta_1}+(L/L_z)^{\beta_2}},
\end{equation}
where the break luminosity, $L_{z}$, evolves with redshift
according to:
\begin{equation}
L_{z}=L_{\star}(1+z)^{\alpha-1}\mbox{exp[$-(z-z_{\star})^2/2\sigma_{\star}^2$]}.
\end{equation} 
In this model, the abundance of quasars peaks at $z_{\star}$ with a
characteristic dispersion of $\sigma_{\star}$. The redshift factor
$(1+z)^{\alpha-1}$ represents the explicit dependence on the spectral
index $\alpha$ where a UV spectrum, $f(\nu)\propto\nu^{-\alpha}$, has
been assumed.   The values of the fitting parameters are
$z_{\star}=2.77$, $\sigma_{\star}=0.91$, $\beta_1=1.83$,
$\beta_2=3.70$, $log (\phi_{\star}/\mbox{Gpc}^{-3})=2.37$, and $log
(L_{\star}/L_{\odot})=13.42$ for an open-universe cosmology with
$h=0.50$, $q_o=0.1$, and $\alpha=1.0$. We use the above values and
convert the LF to our $\Lambda$CDM cosmology by rescaling the
luminosity and volume element for each redshift of interest. Our
conversion of the LF ignores any spectral-dependent $k$-corrections
which are thought to be small assuming an $\alpha=1.0$ power law for
the quasars (Haiman \& Loeb 1998). It is important to point out that
this model predicts that a large fraction of the luminosity at $z>2$
arises from quasars that have not yet been observed.  With a fit to
the LF which goes as $\phi(L)\propto L^{-1.83}$ at the faint end, the
emissivity, $\int \phi(L,z)\ L\ dL$ (in ergs $\mbox{s}^{-1}$
$\mbox{Hz}^{-1}$ $\mbox{cm}^{-3}$), converges only as $L^{0.17}$, and
it becomes apparent that a large portion of the total emissivity will
arise from this regime. Nevertheless, the extrapolation of the LF to
fainter quasars seems to be reasonable based on the analysis of Haardt
\& Madau (1996) where they use the above LF and a realistic form for
quasar spectra to calculate the intensity of the ionizing background,
$J$. Using the cosmological radiative transfer equation for diffuse
radiation (e.g. Peebles 1993),
\begin{equation}
\biggl(\frac{\partial}{\partial
t}-\nu\frac{\dot{a}}{a}\frac{\partial}{\partial
\nu}\biggl)J=-3\frac{\dot{a}}{a}J-c\kappa J+\frac{c}{4\pi}\epsilon,
\end{equation}
where $a$ is the scale factor, $\kappa$ is the continuum absorption
coefficient per unit length along the line of sight, and $\epsilon$ is
the proper space-averaged volume emissivity, Haardt \& Madau include
the effects of absorption and emission by intervening clouds to
calculate the evolution of $J$. They show that $J_{912\ \text{\AA}}$
increases from $\approx 10^{-23}$ ergs $\mbox{s}^{-1}$
$\mbox{cm}^{-2}$ $\mbox{sr}^{-1}$ $\mbox{Hz}^{-1}$ at the present
epoch to $\approx 5\times 10^{-22}$ ergs $\mbox{s}^{-1}$
$\mbox{cm}^{-2}$ $\mbox{sr}^{-1}$ $\mbox{Hz}^{-1}$ at $z=2.5$. This
result is consistent with high-resolution studies of the
proximity effect which give $J_{912\ \text{\AA}}\approx 5 \times
10^{-22}$ obtained for a redshift range $z=1.7-4.1$ (Giallongo et
al. 1996) and $J_{912\ \text{\AA}} \approx 10 \times 10^{-22}$ (Cooke
et al. 1997) for a similar redshift range.  It is furthermore
consistent with limits imposed by the opacity of the Lyman alpha
forest at both high (e.g. Rauch et al. 1997) and low (e.g.  Dav\'e et
al. 1999) redshifts.

While our choice for the quasar LF appears reasonable, one must bear
in mind that it is based on empirical results in only the $B$-band. It
is therefore possible that there may be subtle selection effects which
shroud the underlying form of the LF. In fact, the results of a soft
X-ray {\it ROSAT} survey of active galactic nuclei (AGN) by Miyaji,
Hasinger, \& Schmidt (2000) suggest that the quasar LF differs from
the one presented above. Unlike optically selected quasars, the
evolution of the {\it ROSAT}-selected quasars do not show evidence for
a decrease in number density at redshift higher than 3. In particular,
the X-ray selected AGNs from the survey conducted by Miyaji,
Hasinger,\& Schmidt with Log $L_x>44.5$ have a space density seven
times higher than optically selected quasars from the survey conducted
by Schmidt et al. (1995).  Although the statistical significance of
the difference is marginal ($2\sigma$), the discrepancies may be
indicative of different formation epochs for low and high mass black
holes in the centers of the sources.

To be able to utilize the LF in our selection process, we will first
need to prescribe some way of assigning luminosities to our list of
plausible sources. This will involve adopting a mass-to-light ratio
based on the minimum $B$-band luminosity ($L_{min}$) and a minimum
mass for sources at every epoch.  In Paper I, we considered only the
bright sources and therefore adopted a mass-to-light ratio based on a
constant $L_{min}$ and $M_{min}$. Here we will consider the entire
range of sources responsible for the emissivity implied by the
luminosity function. This requires us to adopt a minimum luminosity
which scales with $z$ as the break luminosity $L_z$ and is chosen to
correspond to a set value $L_{min,o}$ at $z=0$. We assume that the
minimum mass required to form the quasar remains constant with
redshift so that our expression for the mass-to-light ratio becomes,
\begin{equation}
\xi(z_i)=\frac{M_{min}}{L_{min,o}}\mbox{exp[$\frac{z_i^2-2z_iz_{\star}}{2\sigma_{\star}^2}$]}.
\end{equation} 
The assumption that $M_{min}$ is constant in time is motivated by the
idea that there is a certain mass threshold required for the
emission mechanism in quasars to operate.  Both $M_{min}$ and $L_{min,o}$
will act as free parameters in our analysis.

The selection process is initiated by first determining the number
of sources expected at each redshift $z_i$,
\begin{equation}
N_{e}(z_{i})=V_{box}\int_{L_{min}(z_i)}^{\infty}\phi(L_{B},z_{i})\
dL_{B}.
\end{equation}  
where $V_{box}$ represents the comoving volume of the simulation 
and $L_{min}(z_i)=L_{min,o}$
\mbox{exp[$\frac{-z_i^2+2z_iz_{\star}}{2\sigma_{\star}^2}$]}.  $N_{e}$
will thus designate the total number of sources chosen at redshift
$z_{i}$. To convert $N_{e}$ to an integer in an impartial manner, a
random number between $0$ and $1$ is generated and compared to the
fractional component of $N_{e}$. If the fractional component is larger
than the generated number, $N_e$ is rounded up to the nearest integer, else $N_{e}$ is rounded down.

Next, we loop over every plausible source to determine whether it will
be chosen. Our selection criterion will consist of assigning a random
number between $0$ and $1$ to the candidate source of mass $M$ and
selecting it if the assigned value matches or falls below the value of
a some probability function $P(M)$. To ensure that we match the quasar
luminosity function, we define the probability function to be:
\begin{equation}
P(M)=\biggl(\frac{dN}{dM}\biggl)_{LF}\biggl(\frac{dN}{dM}\biggl)_{Hydro}^{-1},
\end{equation} 
where $(dN/dM)_{LF}$ is the expected source mass function for a
simulation box with comoving volume, $V_{box}$, as derived from the
luminosity function:
\begin{equation}
\biggl(\frac{dN}{dM}\biggl)_{LF}=\xi(z_i)^{-1}\phi(M/\xi(z_i),z_{i})V_{box},
\end{equation} 
and $(dN/dM)_{Hydro}$ is just our analytic representation of the mass
function obtained from the actual hydro simulation now rewritten as:
\begin{equation}
\biggl(\frac{dN}{dM}\biggl)_{Hydro}=M^{-1}10^{(a\log{M}+b)}.
\end{equation} 
The loop is terminated when exactly $N_{e}$ sources have been
selected. If the number selected falls short of $N_{e}$ after one run
through the list of plausible sources, the list is reordered randomly
and source selection is continued from the beginning of the list.  In
this manner, a list of sources is compiled for every redshift step of
the simulation, $z_{i}$, with plausible source lists and mass
functions being updated near every hydro output redshift.

Once a source with mass, $M$, has been selected, it is assigned a
$B$-band luminosity, $L_{B}=M/\xi(z_i)$ (in ergs
$\mbox{s}^{-1}$). This luminosity along with an assumed spectral form,
is then used to compute the quantity of He {\small{II}} ionizing flux
that will be generated while the source is active.  In this paper, we
shall assume for all sources, a multi-power-law form for the spectral
energy distribution (SED),
\begin{equation}
L(\nu)\propto
\begin{cases}
\nu^{-0.3} & (2500\ \mbox{\AA}<\lambda<4400\ \mbox{\AA});\\ \nu^{-0.8}
& (1050\ \mbox{\AA}<\lambda<2500\ \mbox{\AA});\\ \nu^{-\alpha_s} &
(\lambda<1050\ \mbox{\AA}),
\end{cases}
\end{equation}
where a choice of $\alpha_s=1.8$ gives the SED proposed by Madau,
Haardt, \& Rees (1999) based on the rest-frame optical and UV spectra
of Francis et al. (1991), Sargent, Steidel, \& Boksenberg (1989), and
the EUV spectra of radio-quiet quasars (Zheng et al. 1998).  We also
allow for the possibility
of bipolar beamed radiation fields by introducing the parameter, $\beta$,
which specifies the beaming angle of the radiation from the sources.
At the end of the entire source selection process, we will have
recorded a list of source locations and ionization rates for every
redshift step of the simulation and we can initiate the radiative
transfer calculations (see Paper I). In selecting the sources and
computing their intensities we have introduced five free parameters
associated with source characteristics, they are: (1) a universal
source lifetime, $T_{life}$, (2) a minimum mass, $M_{min}$, (3) a
minimum luminosity at $z=0$, $L_{min,o}$, (4) an angle specifying the
beaming of the bi-polar radiation, $\beta$, and (5) a tail-end spectral
index, $\alpha_s$, in the regime $\lambda< 1050$ \AA.

In this paper we compute and analyze 6 models with different sets of
values for the free parameters. Table 1 lists the choices for each
model.  Model 1 will represent our fiducial case. In this model, the
value for the source lifetime simply reflects a reasonable choice
based on the quasar light curve derived by Haiman \& Loeb (1998) for a
similar cosmology and luminosity function. Our choice for $L_{min,o}$
in the model is based on the results of Cheng et al. (1985) who show
that the LF of Seyfert galaxies (which are well correlated with that
of optically selected quasars at $M_B=-23$) shows some evidence of
leveling off by $M_B\simeq -18.5$ or $L_{min}\simeq 6.44\times 10^{9}
\ L_{B,\odot}$ at $z=0$. The value of $M_{min}$ is then chosen such
that the $B$-band emissivity from the selected sources matches the
expected emissivity derived from the LF. Given our value for
$L_{min,o}$, we find $M_{min}=1.42 \times 10^{10}\ M_{\odot}$ is able
to match the derived $B$-band emissivity to within a few percent at
z=3 for this particular case. This value seems reasonable given the
assumption that the sources are galaxy type objects acting as hosts
for quasars. It is also important to note that the underlying
cosmological simulation used in this paper provides an acceptable
level of resolution at this mass limit and thereby provides us with a
realistic mass function for the source selection process. Furthermore,
we are able to reproduce roughly the same range of observed $B$-band
luminosities for the brightest quasars in the survey conducted by
Warren, Hewett \& Osmer (1994). Consequently, our ability to properly
match the high end of the luminosity function with our most massive
groups ensures that our volume is large enough to properly sample the
range of source masses. The next parameter in the model describes the
way sources beam their radiation. Here in our fiducial model, we
choose our sources to radiate their flux isotropically by setting the
beaming angle of the bi-polar radiation field to $\beta=\pi$. And for
the final parameter, the tail-end spectral index, we choose the value
$\alpha=1.8$, making the SED in this model identical to the one
proposed by Madau, Haardt, \& Rees (1999).

Model 2 is identical to model 1 except for the anisotropic casting of
the radiation identified by a beaming angle of $\pi/2$. In models 3
and 4, we retain all the values as in our our fiducial model (model 1)
but vary the source lifetimes by a factor of 2 in each direction. In
model 5, in an attempt to examine the effect of a reduced level of
ionizing emissivity, $L_{min,o}$ is increased by a factor 4 and the
spectral index is set to $\alpha=2$. For model 6, we also reduce the
expected ionizing emissivity, however we do so only by steepening the
spectral index by itself to the value of $\alpha=2.3$. 

It is important to point out that even the brightest and most long
lived sources arising in any of the models have ionizing intensities
which limit them to spheres of influence that are much smaller
than the size of the simulation box. As a result, our simulation box is
sufficiently large to statistically examine the epoch when
sources are just turning on.  When the medium has become mostly
ionized, we utilize enough rays to properly track the contribution of
ionizing photons from each source to the diffuse component of the
radiation field.

The six models described above each predict a unique set of sources
which are responsible for the ionization of the IGM. To compare the
differences, we ran each model on the same cosmologically evolving
density field (see \S 4.3 of Paper I for details regarding the
evolution of the grids) from $z\simeq5.7$ to $z\simeq2.7$. We discuss
the results from these models in the next section.

\section{RESULTS}

In this section, we discuss the outcome of the simulations and present
an analysis geared towards identifying and differentiating global
characteristics of the models. In addition, we will describe our
method for extracting artificial spectra from the outputs and present
a comprehensive statistical study of spectral features extracted from
each model.

\subsection{Ionizing photons}

In Figure 1 we show the number of He {\small II} ionizing photons per
comoving volume released from the sources as function of redshift for
each model. The solid histogram in each plot specifies the number of
ionizing photons released per unit volume at each redshift bin and the
dashed curve represents the corresponding cumulative value. The widths
of the bins in each histogram represent the source lifetime in the
model.  Models 1, 3, and 4, differ only in the source lifetime
parameter; consequently, the total number of sources invoked in these
models is in direct proportion to the ratios of their source
lifetimes: 2:4:1, respectively. Model 2 has a greater abundance of
sources relative to model 1 due to the larger number of sources needed
to match the implied number density from the observed LF, assuming the
radiation from the sources is being beamed into two symmetric polar
cones of width $\beta$ radians. In particular, assuming the polar
directions for the radiation are randomly distributed (a condition
which we have implemented), the number of expected sources is
proportional to $(1-\cos{\frac{\beta}{2}})^{-1}$. This amounts to an
excess factor of $3.41$ for the number of sources in model 2 relative
to model 1. Note, however, that the number of ionizing photons
released in both models is the same. In fact, this is true for models
1-4 where we have matched the values for the parameters $M_{min}$,
$L_{min,o}$ and $\alpha_s$.  In models 5 and 6, the new choices for
$L_{min,o}$ and $\alpha_s$ lead to a relative paucity in the number of
ionizing photons compared to models 1-4 (see Figure 1). In particular,
the cumulative number of ionizing photons released in the $10^6 \
\mbox{Mpc}^3$ comoving volume by $z\simeq2.75$ in models 1-4 was
$4.08\times 10^{72}$ compared to $1.34\times 10^{72}$ and $1.47\times
10^{72}$ for models 5 and 6 respectively. It is thus clear that the
choices for $L_{min,o}$ and $\alpha_s$ have a substantial influence on
the amount of ionizing radiation that is emitted from the sources.
This is also true for the parameter $M_{min}$, but we have chosen to
hold it fixed in the models considered in this paper.

To illustrate how the results in Figure 1 depend on each parameter, in
Figure 2 we compare the cumulative number of ionizing photons that are
released per unit comoving volume as a function of redshift between
models as the parameters are varied separately. It is evident that
there is a strong dependence on each of the parameters and this will
be an important point to consider when discussing the evolution of the
reionization process and its sensitivity to the cumulative influence
of ionizing radiation.

\subsection{Global ionization fractions}

We now discuss global properties of the ionization state of the volume
for the various models considered.  In Figures 3a and 3b we show how
the ionized mass and volume fraction, respectively, evolve with
redshift.  In all our models, He {\small II} is mostly fully ionized by
$z\simeq 3.3$, a result which is in accord with observations (Jakobsen
et al. 1994; Davidsen, Kriss, \& Zheng 1996; Hogan, Anderson, \&
Rugers 1997; Reimers et al. 1997; Anderson et al. 1999; Heap et
al. 2000; Smette et al. 2000).  The evolution to a completely ionized
state is quite similar for all the models with the exception of models
5 and 6 where the lower emissivities lead to a more gradual rise in
the ionized fractions.  Nevertheless, the entire volume is essentially
fully ionized by $z\approx 3.3 - 3.8$ in all our models.  Thus, we
predict that the opacity of the He {\small II} Lyman alpha forest
should increase rapidly for $z > 3$ and that the redshift evolution of
this opacity will contain information that can be used to discriminate
between models like those considered here (i.e. models 5 \& 6 versus
models 1-4).

It is also quite evident from Figures 3a and 3b, that whereas the
volume is essentially fully ionized by $z\sim 3.3$, this is not
necessarily true of the mass.  This is a consequence of extremely
overdense cells with large clumping factors which quickly recombine
after they have been ionized. In models 5 and 6, lower photoionization
rates make it even harder to keep these cells ionized.  Due to their
large overdensities (factors of $10^3$ or more) it takes relatively
few such cells to substantially lower the ionized mass fraction.
These cells will not, however, strongly influence the mean opacity of
the He {\small II} Lyman alpha forest, which directly measures the
volume fraction ionized. 

In Figure 3c, we plot the ratio of the ionized mass fraction to the
ionized volume fraction as a function of redshift for each model.
Here, we can more clearly see how the ionized mass fraction dominates
the corresponding volume fraction at high redshifts. This is
consistent with the idea that at earlier times, when the medium is
mostly neutral, the densest and most compact regions immediately
surrounding the sources are the first to become ionized.  However,
with time, the ionization fronts expand into the less dense but more
voluminous voids, thereby decreasing the ratio shown in Figure 3c and
eventually leveling it off to a value slightly less than unity.

One may conjecture that the epoch of overlap, when intervening patches
of He {\small II} no longer obscure the field of view from any source
and the universe becomes transparent on a much larger scale than
during the previous stage, may be reached as the ionized volume
fraction approaches unity.  Testing this idea requires additional
analysis into the evolution of the ionizing background. Namely, once
the volume is sufficiently ionized so that every sources can {\it see}
every other source, the ionizing intensity of the background is
expected to rise sharply as the contributions from 
different sources combine.
Such an effect was reported by Gnedin (2000), whose simulations of
cosmological reionization by stellar sources predict an overlap
epoch in hydrogen that is characterized by a sharp rise in the level
of the ionizing background and in the photon mean free path.

To see if we predict a similar rise, we have plotted in Figure 4 the
number of ionizing photons per helium atom available in the diffuse
component of the radiation field as function redshift for model 1
(models 2-4 demonstrated very similar results and are not shown). Here
we do indeed see a similar steep rise in the background with
decreasing redshift starting at $z\simeq3.7$.  From Figure 3b, we note
that this is right around the time when the corresponding volume
fraction for model 1 has reached unity. There is a rise of nearly a
factor of 100 in the background intensity
between $3.2\leq z \leq3.7$ (corresponding to the region
between the dashed lines) before leveling off. It thus seems
reasonable to label this short era as the epoch of overlap immediately
preceding complete reionization, which in this case would 
be reached at
$z\simeq3.2$. In the case of models 5 and 6, the volume fraction does
not reach unity until $z\sim3$, near the redshift where we stop our
calculations. We thus would expect a similar type of rise in the
corresponding backgrounds in these models as well. However, we
speculate whether the rise would be as dramatic as the one seen in
model 1 since in models 5 and 6 the epoch of overlap occurs right
around the time when the LF predicts a steep downturn in the ionizing
emissivity of the contributing sources.

\subsection{Clumping characteristics within ionized regions}

To track how I-fronts expand into the densest structures, we track the
evolution of the mean clumping factor in ionized regions. In Figure 5,
we plot the evolution of the ratio of the clumping factor of the
ionized gas to the clumping factor of the entire volume as a function
of redshift for each model.  Clumping factors are computed on a fixed
mesh of limited resolution based on an underlying smoothed density
field. As such, one should be careful not to attach physical
significance to the actual values, but rather view them as a tool to
differentiate between the morphological evolution in each model.  Our
condition for considering a cell to be {\it ionized} is that the gas
within is ionized by at least $90\%$ in mass.  An immediate
consequence of this is that models 5 and 6, which lack the emissivity
to retain high ionization fractions in the densest cells, have much
lower clumping factors within their {\it ionized} regions.  In models
1--4, the systematic increase in the clumping factors once most of the
volume has been ionized ($z\lesssim 3.3$) is consistent with the
interpretation presented by Miralda-Escud$\acute{\text{e}}$, Haehnelt,
\& Rees (2000) in which they argue that in the post-reionization
stage, the densest clouds and filaments (which dominate the overall
clumping factor) gradually become ionized as the mean ionizing
intensity increases. Nevertheless, a general feature among all the
models is the tendency of the clumping factor of the ionized regions
to remain well below the levels of the overall clumping factor for the
entire volume, even at $z\simeq 2.8$. This feature is a consequence of
a small number of highly overdense, mostly neutral cells which
dominate the overall clumping factor. The difficulty in ionizing these
cells is also reflected in the discrepancy between the mass and volume
ionization fractions in Figure 3.

\subsection{Photoionization rates}

In the left panel of Figure 6, we plot the volume-weighted global He
{\small II} photoionization rate as a function of redshift for each
model. Global rates were obtained at discrete redshifts by averaging
over the rates in each cell of the corresponding grid (see \S 3.3 of
Paper I for a description on how photoionization rates are
computed). Note that the rates begin to level off at $z\sim3$,
consistent with the LF used in this paper. The plot also shows why
photoionization proceeds more slowly in models 5 and 6 compared with
the others.  In addition, the radiation field in model 3, although
not as weak as those in models 5 and 6, is somewhat weaker than that
in models 1, 2, and 4.

In the right panel of Figure 6, we plot the ratio of the dispersion in
the photoionization rates to the corresponding photoionization
rate. The dispersion in the rates is remarkably high and explicitly
demonstrates the non-uniform nature of the radiation field from local
sources. The particularly large amplitude of the dispersion in models
5 and 6 relative to the others is a natural consequence of the higher
opacity in these models which inhibits uniformity in the
photoionization rate.  It is also interesting to point out the
somewhat larger dispersion seen in models 2 and 3 relative to models 1
and 4.  This is most likely due to the large number of sources in
these models which can lead to a highly irregular radiation field. In
the case of model 2, there is the additional ingredient that the
ionizing flux is released anisotropically ($\beta=\pi/2$) from the
sources.  The cone shaped beaming of the radiation in this model only
enhances the other factors responsible for large dispersions in the
photoionization rates.

\subsection{Spectral analysis}

It is now believed that absorption by diffuse, cosmologically
distributed gas is responsible for the hydrogen Lyman alpha forest
(e.g. Cen et al. 1994; Zhang et al. 1995; Hernquist et al.  1996).
Similarly, Ly$\alpha$ absorption by He {\small II} along a line of
sight (LOS) to a distant quasar probes gas in the intervening IGM at
even lower overdensities (Croft et al. 1997).

To relate our models to observations of quasars, we extract artificial
Ly$\alpha$ absorption spectra from the simulation outputs and then
derive statistical properties of the spectral features seen in each
model.  Of particular importance is the He {\small II} mean optical
depth, which we compare to observational results in \S 4.5.4.

We extract artificial Ly$\alpha$ absorption spectra using the particle
information in the SPH simulation coupled with the time-dependent
ionization fractions from the radiative transfer calculations. For
each model, we generate 500 spectra along randomly selected lines of
sight between $z\simeq2.8-3.6$.  Our procedure is similar to that
contained in the TIPSY software package (Katz \& Quinn 1995), but does
not require that a LOS be perpendicular to a box face in the
simulation volume.  Each LOS has a unique and arbitrary direction
relative to the box coordinate system and wraps through the simulation
volume repeatedly via periodic boundaries.  Our method also
interpolates the hydrodynamic quantities and ionization state
variables between consecutive data dumps and hence can construct
spectra of any length in redshift space.

Using the smoothing kernels of the SPH particles, gas densities and
temperatures are computed along a LOS at intervals corresponding to a
resolution of at least $\lambda=0.03$\ \AA. The component of the
peculiar velocity of the gas in the direction of the LOS is also
computed at each point.  Once all physical quantities have been
gathered, Voigt profiles are fitted to each spectrum by interpolating
between the corresponding line-absorption coefficients provided in
Harris (1948). It is important to point out that in computing the line
profiles we use a minimum gas temperature of $2.0\times10^4 \ K$ as a
correction to the SPH temperatures which exclude the extra heating
introduced by radiative transfer effects (see Abel and Haehnelt,
1999).

Information related to the hydrodynamical state of the particles is
updated at SPH output redshifts of $z=$ 3.59, 3.47, 3.36, 3.24, 3.14,
3.03, 2.94, 2.84 and 2.75. Ionization fractions, which evolve on
shorter timescales, are updated after each source lifetime by mapping
grid outputs from the radiative transfer calculations onto the
particles. Moreover, the arrival times of the I-fronts (see section \S
3.1.3 of Paper I) in each grid are used to discriminately update the
fractions along the LOS within a single source lifetime interval.

We have tested our technique for extracting physical properties as a
function of redshift against the results of the TIPSY software package
for lines of sight perpendicular to box faces of the simulation volume
and find excellent agreement in all cases.

Figure 7 shows a sample spectrum from model 1 along a LOS that spans
$z\simeq2.8-3.6$ together with corresponding plots of the singly ionized
fraction and gas density as a function of redshift. The rise of the
neutral fraction with redshift and the corresponding decrement in
transmission is quite evident from the panels in Figure 7.

In Figure 8, we plot transmission spectra for the different models
along a particular LOS.  In models 1, 2 and 4, where most of the IGM
is in a highly ionized state, one can easily recognize similarities in
the spectral features due to the underlying density and velocity
structures.  Models 3, 5 and 6 are different in this respect, because
the relatively large neutral fractions in the gas lead to high optical
depths which shroud the underlying structures. This is especially true
for models 5 and 6 where large sections of the spectra near the red
end are completely opaque due to weak ionizing intensities.

The dashed vertical line through the panels in Figure 8 at 1307 \AA \
is a crude attempt to separate the spectra into two regimes that
represent a high ({\it left}) and low ({\it right}) level of
ionization in the gas based on the general features in models 1-4. We
use this dividing point in our forthcoming analysis of the spectral
features in an effort to probe the evolution of the ionization state
of the IGM using absorption spectra.

We note that the spectra extracted from models 5 and 6, where the
volume fraction of ionized He {\small II} starts to drop rapidly for
$z > 3.3$ (see Figure 3), exhibit extended wavelength intervals with
virtually no transmission.  A similar behavior is seen in models 1-4
for $z > 3.8$.  Thus, our analysis predicts that future observations
of the He {\small II} Lyman alpha forest in quasar spectra should show
a dramatic drop in transmission during the epoch of He {\small II}
reionization, between $z \sim 3.3 - 3.8$.

\subsubsection{Intersections versus transmittance}

In this section we consider a more comprehensive analysis tool that
utilizes information from the spectra at all transmission levels. The
analysis involves tracking the number of intersections that occur
through the spectra as a function of transmittance. Figure 9 shows
the results of the analysis for each model averaged over 500 lines of
sight, again focusing on two separate spectral segments. In both
segments we see dramatically different results between models.  In
segment B ({\it left}), which represents the high redshift component
of the spectral range, we note that the curves from models 1, 2 and 4
appear to posses local maxima near the 30\% transmittance level.  The
curves from models 3, 5 and 6 exhibit no such peak and instead possess
exponential type forms. In segment A ({\it right}) we notice that the
curves from model 1, 2 and 4 have now converged to a skewed form which
peaks near a transmittance of 80\%. Furthermore, model 3 appears to
have developed a local maximum with a form similar to the curves seen
in models 1, 2 and 4 from segment B.  Interestingly, models 5 and 6
appear to be following in model 3's footsteps.

A strong evolution of spectral features is clearly seen in these
plots. In particular, in the models where the IGM has not yet become
thoroughly ionized, the intersections decrease asymptotically with
transmittance. However, as the IGM is gradually ionized to higher
degrees, the intersections appear to peak first at low transmittance
levels then move towards higher levels. Eventually, the curves
converge to a form that rises linearly with transmittance, peaks, and
then asymptotically falls off near a transmittance of 100\%.

It is clear from this discussion that even as simple a diagnostic tool
as that shown in Figure 9 can strongly discriminate between models
and their rates of evolution.  For example, consider model 3 whose
spectrum in segment A visually looks very much like the spectra from
models 1, 2 and 4 (see Figure 8), yet in the right-hand panel of
Figure 9, one can clearly see the disparity.  On this basis, we
anticipate that similar tools applied to observations of the He
{\small II} Lyman alpha forest forest should constrain the nature of
the sources responsible for the hard component of the ionizing
radiation field.

\subsubsection{Probability distribution function of transmittance}

We next consider the probability distribution function of the
transmittance (TPDF) as another tool to study the properties of the
absorption spectra.  The TPDF for each model was measured by averaging
over 500 spectra, with all pixels weighted equally. We use 30 bins of
equal width between the 0 and 1 transmission level. Figure 10 shows
the TPDF ({\it left}) and the cumulative TPDF {\it right} for each
model.  Again, there is an obvious contrast between the models. The
TPDF for models 1, 2 and 4 exhibit strong peaks near a transmittance
of 0.9 while models 5 and 6 clearly do not; Model 3, however, seems to
again reflect an intermediate case, possessing a peak near 0.9, but
not at an amplitude similar to those of models 1, 2 and 4. The sharp
rise near zero transmittance seen in all the models reflects the presence
of trough-like features in the spectra. Naturally, as the IGM becomes
more transparent, such features become less prevalent. This attrition
is made evident from the relatively small amplitudes seen in the
cumulative TPDF near zero transmittance for models 1, 2, and 4, all of
which harbor gas with relatively high ionization fractions.

The form of the TPDF implied by our models is reminiscent of the
transmitted flux PDFs (TFPDF) measured for the hydrogen Lyman alpha
forest by McDonald et al. (2000). In particular, they use a sample of
eight quasars observed at high redshift to determine the TFPDF in
three redshift bins centered on $z=$2.41, 3.00, and 3.89.  In their
highest redshift bin ($z=$3.89), the TFPDF measured by McDonald et al.
resembles the TPDF from our model 3.  At their lower redshifts
($z=$3.0, 2.41), their TFPDF's begin to resemble the results from our
models 1, 2 and 4.

\subsubsection{Redshift evolution of the mean optical depth}

Figure 11 shows the redshift evolution of the effective mean optical
depth for He {\small II} absorption, which we define as
$\bar{\tau}_{\mbox{\tiny{He II}}}\equiv-\log_e \langle T \rangle$,
where $T$ is the transmittance. The average is performed over 500
lines of sight within 35 wavelength bins of width $\Delta
\lambda=6.57$ \AA. Hatched regions represent the optical depth derived
from the simulations at the 95\% confidence level with the dashed
lines indicating mean values. For comparison, we also plot the
opacities measured at different redshifts in the spectra of Q 0302-003
(Heap et al. 2000), PKS 1935-692 (Anderson et al. 1999, reported
values come from Smette et al. 2000 who perform an optimal reduction
of the whole data set), and HE 2347-4342 (Smette et al. 2000).
Horizontal lines through the data points represent the redshift range
over which the opacity was averaged. Data points with no horizontal
lines represent measurements made at specific wavelengths and should
be assigned less weight as they do not represent mean opacities.

For models 1, 2, and 4, it appears that most of the measured opacities
are much larger than predicted by our analysis.  Model 3 does better,
though falls short of the majority of data points near $z \simeq
2.85$. On the other hand, models 5 and 6 exhibit larger optical
depths, and model 5, in particular, appears to be in 
reasonable agreement
with the data. It is important to point out, however, that the
observed values are likely subject to a great deal of statistical
variance.  In particular, the hatched regions in Figure 11 represent
averages over 500 lines of sight, whereas the observational data stem
from only 3 lines of sight. One therefore needs to be careful when
making judgments regarding the {\it best-fit} model.  Nevertheless,
such comparisons provide an important point of reference upon which to
base claims regarding the nature of the sources. We shall discuss such
conclusions in the following section.

\section{DEPENDANCE ON CLUMPING AND TEMPERATURE OF THE IGM}

Both the temperature and clumping factor of the IGM play an important
role in the evolution of the reionization process by influencing the
rate at which recombinations take place in ionized regions. The
analysis presented in this paper uses approximate techniques to set
the values of these parameters; it is therefore necessary to explore
the associated parameter space to acquire some intuition for their
potential impact on the reionization process.

In Figure 12, we compare ionized mass ({\it left}) and volume ({\it
right}) fractions between our fiducial model 1 and two additional
models, model 1a and 1b, which are identical to model 1 except for
temperatures and clumping factors. In model 1a, the temperature of the
ionized gas was set to $7000 \ K$ as opposed to $2.0\times10^4\ K$ as
in model 1. In model 1b, all the information regarding the clustering
of particles was disregarded by setting cell clumping factors to
unity.  We note that these are extreme limiting cases to consider, but
they do indicate potential sources of error in our analysis.

In the case of model 1a, the slight drop in the ionized fractions
reflects the greater difficulty involved in ionizing a relatively
cooler medium where recombinations are more prevalent. This dependence
of the optical depth on the temperature of the gas is also well known
in the case of the hydrogen Lyman alpha forest (see Machacek, et
al. 2000 for a detailed discussion and references).  Conversely, in
model 1b, recombinations are less prevalent than in model 1 due to the
adoption of a uniform distribution of the gas in each cell ($C_f=1$)
and as a result ionization fractions are consistently larger at each
redshift.  More specifically, at $z=3$, models 1a and 1b produce
respectively a -3.3\% and 4.7\% difference in the ionized mass fraction
compared to model 1.

A comparison of the mean optical depths between these cases is shown
in Figure 13. The results were again computed using a statistical pool
of 500 lines of sight. The relative opacities seen in model 1a and 1b
are consistent with the discussion above.  The shift in temperatures
between models 1 and 1a appears to have a more significant impact on
the mean optical depth than does ignoring the clumping of gas within
cells.  However, it does not appear that temperature variations can
reconcile the deficit between the observed values of the mean optical
depth and those inferred from our analysis for this particular model.

\section{SUMMARY AND CONCLUSIONS}

The most important factor determining the reionization history of He
{\small II} is the cumulative number of ionizing photons that are
produced by the sources.  Given a specific form for the LF, ionizing
emissivities are ultimately determined by intrinsic source
characteristics such as their minimum luminosity and spectral energy
distribution. In the analysis presented here, a numerical method
designed to follow the inhomogeneous reionization of a cosmological
volume by a set of point sources (quasars) was used to explore a small
portion of the parameter space associated with these
characteristics. Six models describing different sets of source
properties were examined.  Models 1-4 differ in source lifetimes and
radiation beaming angles, but were all fixed to produce the roughly
the same number of ionizing photons. More specifically, the tail-end
spectral index of the SED in these models was set to $\alpha_s=1.8$,
as in the SED proposed by Madau, Haardt, \& Rees (1999), while the
minimum B-band luminosity at $z=0$ was set to $L_{min,o}=1.42 \times
10^{9} \ L_{B,\odot}$ based upon an estimate from the results of Cheng
et al. (1985).  For comparison, models 5 and 6, were specifically
designed to the produce substantially less ionizing emissivity while
retaining plausible values for $\alpha_s$ and $L_{min,o}$ (see Table
1). As expected, there is a sharp contrast between the results from
the two groups.  Specifically, the IGM in models 1-4 is ionized
earlier and more thoroughly than in models 5 and 6, resulting in
significantly lower mean opacities for the spectra (see Figure 3 and
11) at redshifts $z\sim 3$.

Based solely on the limited number of measured He {\small II}
opacities from observed quasar spectra, it appears that the ionizing
emissivities in models 1-4 are too high, resulting in mean opacities
lower than the observed values at all redshifts.  If we accept our
choice for $M_{min}$ as being reasonable, then the SED and/or
$L_{min,o}$ need to be adjusted in order to better match the
observations.  In model 5, we picked a value of for $L_{min,o}$ that
was a factor of 4 times higher than the value in models 1-4.  Although
this seems like a large difference from the original estimate of Cheng
et al. (1985), a quick glance at Figure 1 of their paper reassures us
that such a choice for the limiting luminosity (magnitude) for quasars
is still plausible if we assume that the LF of Seyfert galaxies is
similar to the LF of quasars.  The tail-end spectral indices in models 5
and 6 were set to $\alpha_s=2.0$ and $\alpha_s=2.3$ respectively,
which also represent plausible choices given the range of indices
derived from observations of the EUV spectra of radio-quiet quasars at
intermediate redshifts (Zheng et al. 1998).

That model 5 provides the best match to observations of the mean
optical depth bolsters the corresponding choices for $\alpha_s$ and
$L_{min,o}$.  However, as model 6 makes clear, varying $\alpha_s$
alone within a plausible range yields more than enough leeway to also
match the observations while retaining the more widely quoted value
for $L_{min,o}$ (see, e.g., Haardt \& Madau 1996; Madau, Haardt, \&
Rees 1999; Bianchi et al. 2001). Furthermore, the results from model 3
suggest that the choice for source lifetimes may also affect the
overall ionization state of the IGM, adding a further free parameter.

However, uncertainties in the clumping, because of small scale power,
and the gas temperatures do not allow us to clearly eliminate specific
models using existing observational data. Therefore, we refrain from
discussing best fits to the observations since we clearly have not
conducted a comprehensive study of the associated parameter
space. Rather, the main purpose of this paper is to provide a basic
overview on how source characteristics influence global properties of
the reionization process.

Some specific insights from our analysis are as follows:

1. The systematic increase in the clumping factors associated with
ionized regions towards low redshifts once most of the volume has
already been ionized is consistent with the argument that in the
post-reionization stage, the densest clouds and filaments become
gradually ionized to higher degrees as the mean ionizing intensity
increases.

2.  The dispersion of the mean photoionization rates appears to be
remarkably large and explicitly demonstrates the non-uniform nature of
a radiation field dominated by local sources. Although dispersion
levels do show signs of decreasing at the lowest redshifts when
background intensities begin to exceed the intensities from local
sources, the results demonstrate the importance of solving the
radiation field locally during and before the epoch of
reionization. The relatively large dispersion seen in models 2 and 3
can be attributed to the large number of sources which can lead to
highly irregular radiation fields.

3. Measurements of the number of intersections as function of
transmittance and the TPDF of Ly$\alpha$ spectra offer insightful
information regarding the ionization state of the IGM at specific
times.  In particular, both methods appear to be very sensitive to
small differences in the underlying ionization fractions and as a
result serve as useful tools to differentiate between spectra (models)
that at first glance seem quite similar.

4. The lifetimes of the sources responsible for reionization may have
an impact on the overall evolution of the process. Since sources tend
to live in the densest regions of the IGM, the ionizing flux must
first break through their dense surroundings in order to reach the
IGM. Such an effect has been postulated for the spectrum of
HS1700+6416 (Davidsen et al. 1996).  One may therefore argue that the
inability of model 3 to bring the IGM to high level of ionization
relative to models 1, 2 and 4 might have to do with the fact that the
relatively short lifetime requires a larger number of sources to be
involved in reionization, effectively requiring more of the total
ionizing flux to emanate from inside the densest structures. As a
result, the IGM may be subject to less of the cumulative ionizing 
potential
of the sources. One might wonder why model 2, which actually invokes
even more sources than model 3 does not yield similar trends. In
that case, the beaming of the radiation allows the sources to more
easily "break out" of the dense absorbers which surround them thereby
allowing more of the radiation to reach the diffuse IGM. The opposite
case to model 3 occurs for model 4 where the longer lifetimes require
fewer sources be invoked. In this case, by definition, there will be
relatively fewer absorbers through which the sources have to shine
their flux during the course of reionization. Consequently, the IGM in
model 4 appears to be relatively more ionized than the other models.

Our analysis reflects a first step to model radiative transfer effects
and to develop new insights into open questions regarding the
reionization of the universe. In particular, the method used here can
be modified to study the reionization of H {\small I}, offering
further opportunities for comparisons with observations, especially in
light of recent evidence for the reionization of H {\small I} at
$z\sim6$ (Becker et al. 2001). Furthermore, analysis of both H {\small
I} and He {\small II} Ly$\alpha$ opacities from simulated spectra and
observational results can constrain the contribution of starburst
galaxies to the UV ionizing background (see, e.g., Bianchi et
al. 2001; Kriss et al. 2001). We also envision supplementing our
method to include the effect of heating from photoionization to
explore the possibility of searching for fluctuations in the IGM
temperature using the Ly$\alpha$ forest (see, e.g, Zaldarriaga
2001).

\begin{acknowledgments}
We thank Scott Ransom for his assistance as system administrator of
the computer cluster which was used to carry out the calculations. We
also thank Abraham Loeb and Chris Metzler for informative discussions
regarding the specifics of luminosity functions. We are also very
grateful for Nick Gnedin's enlightening commentary regarding the
technical aspects of our method.  This work was supported in part by
NSF grants ACI96-19019, AST-9803137, and PHY 9507695. 
\end{acknowledgments}

\clearpage

\begin{table}[htb]
\begin{center}
\begin{tabular} {cccccc}
\multicolumn{6}{c}{\textbf{TABLE 1 }} \\
\multicolumn{6}{c}{Source Parameters } \\
\hline
\hline
Model  & $T_{life}$ [$10^7$ yrs] & $L_{min,o}$ [$10^{9} L_{B,\odot}$]  &$M_{min}$ [$10^{10} M_{\odot}$] & $\beta$ [radians] & $\alpha_s$ \\
\hline
1 & 2.0 & 6.44 & 1.42 & $\pi$ & 1.8\\
2 & 2.0 & 6.44 & 1.42 & $\pi/2$ & 1.8 \\
3 & 1.0 & 6.44 & 1.42 & $\pi$ & 1.8\\
4 & 4.0 & 6.44 & 1.42 & $\pi$ & 1.8\\
5 & 2.0 & 25.8 & 1.42 & $\pi$ & 2.0\\
6 & 2.0 & 6.44 & 1.42 & $\pi$ & 2.3\\
\hline
\end{tabular}
\end{center}
\end{table}

\clearpage

\begin{figure}[htb]
\figurenum{1}
\setlength{\unitlength}{1in}
\begin{picture}(6,6.5)
\put(-0.65,-1.7){\includegraphics{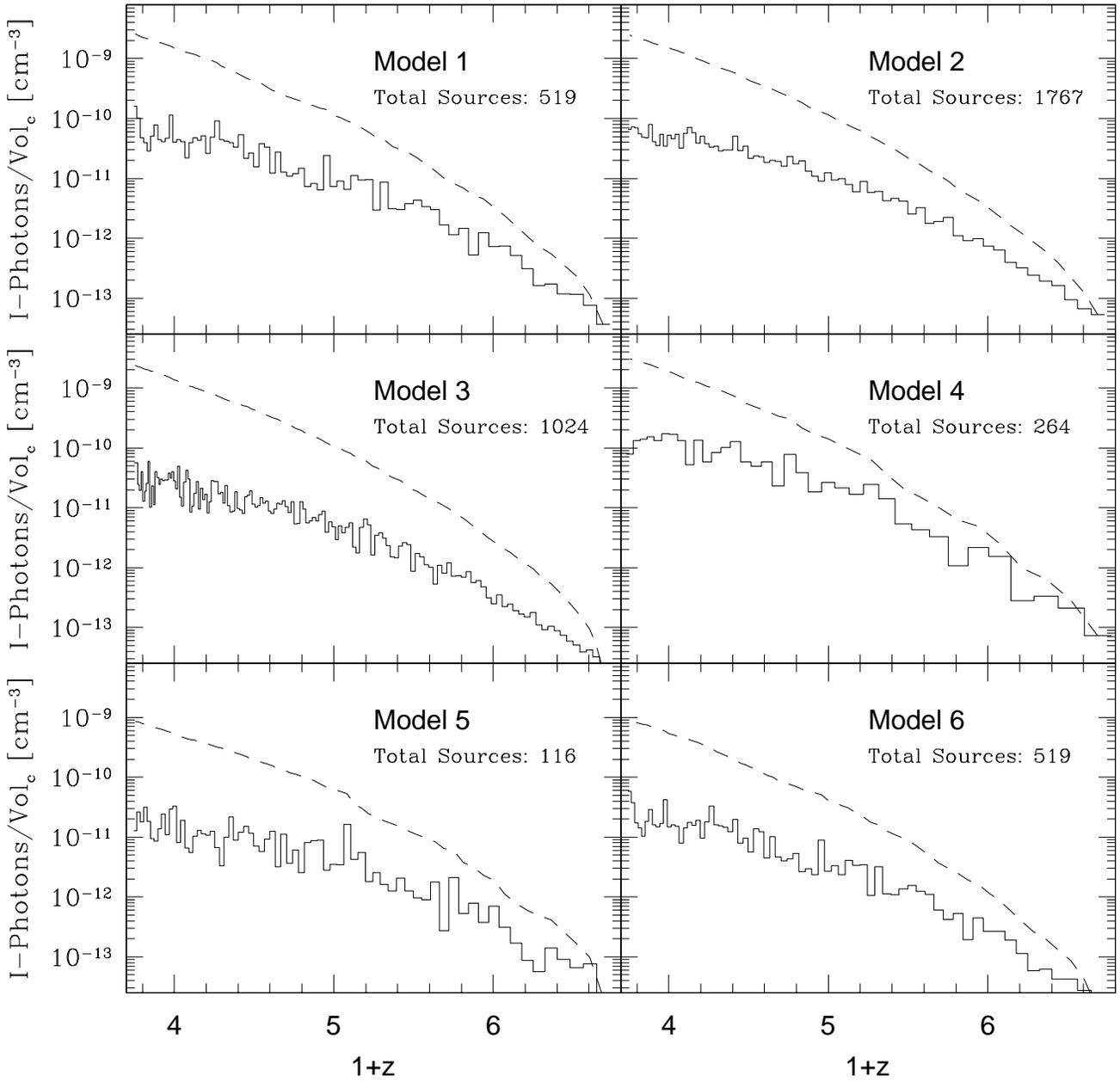}}
\end{picture}
\caption{ The number of He {\small II} ionizing photons per unit comoving
volume released from the sources as a function of redshift in each of
the six models. The solid histogram in each plot specifies the number
of ionizing photons released per unit volume in each redshift bin and the
corresponding dashed curve represents the cumulative value. The widths
of the bins in each histogram represent the associated source
lifetimes. The total number of sources invoked is noted in the upper
right of each plot (see text for discussion).  }
\end{figure}

\clearpage

\begin{figure}[htb]
\figurenum{2}
\setlength{\unitlength}{1in}
\begin{picture}(6,6.4)
\put(-0.65,-1.8){\includegraphics{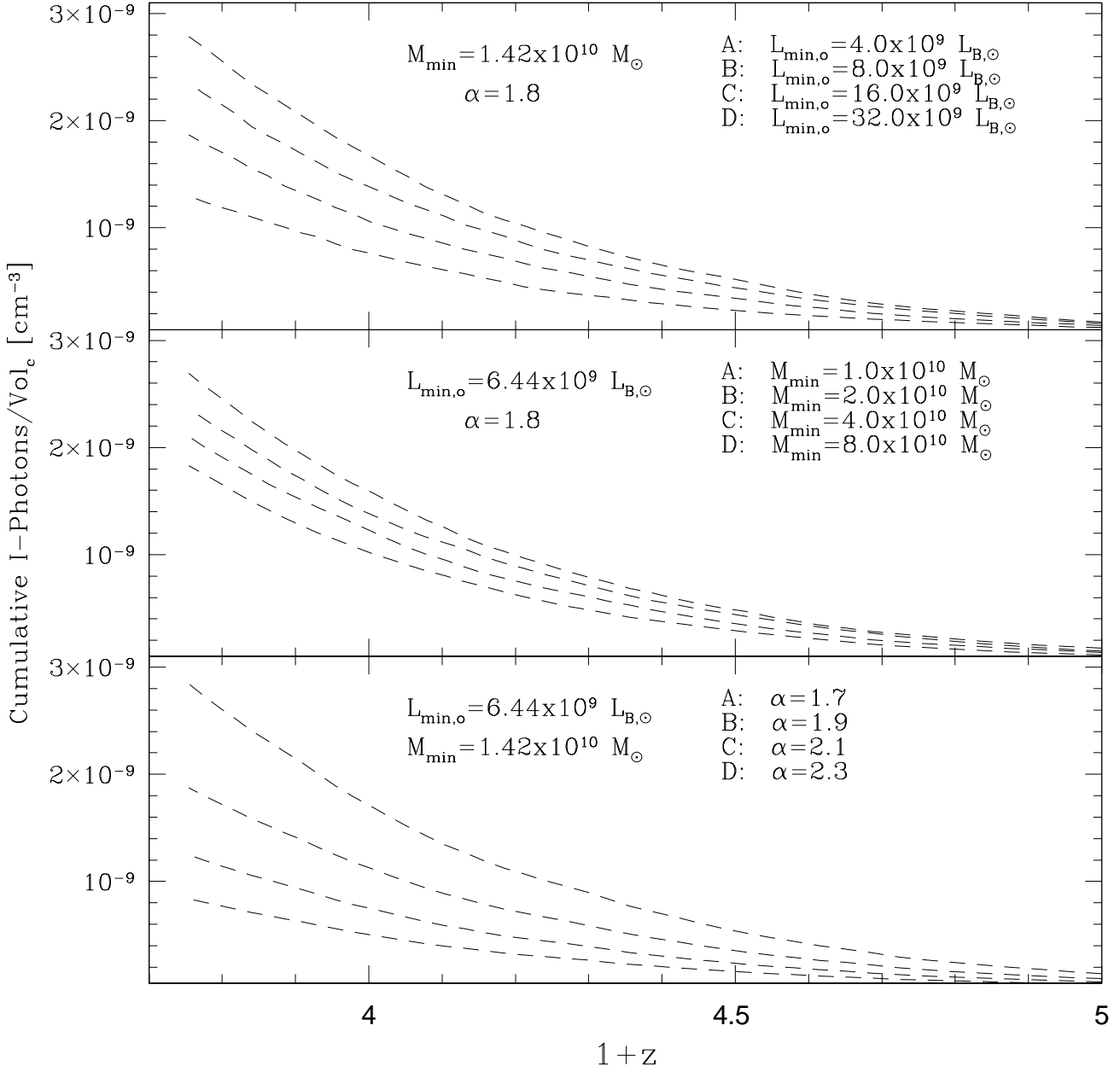}}
\end{picture}
\caption{Cumulative number of ionizing photons that
are released per unit comoving volume as a function of redshift in our
analysis as we vary the parameters $M_{min}$, $L_{min,o}$, and
$\alpha_s$ separately (see text for definitions). Labels A, B, C, and
D refer to the curves in each plot in a top-down respective order.}
\end{figure}

\clearpage

\begin{figure}[htb]
\figurenum{3}
\setlength{\unitlength}{1in}
\begin{picture}(6,6.5)
\put(-2.35,-1.65){\includegraphics{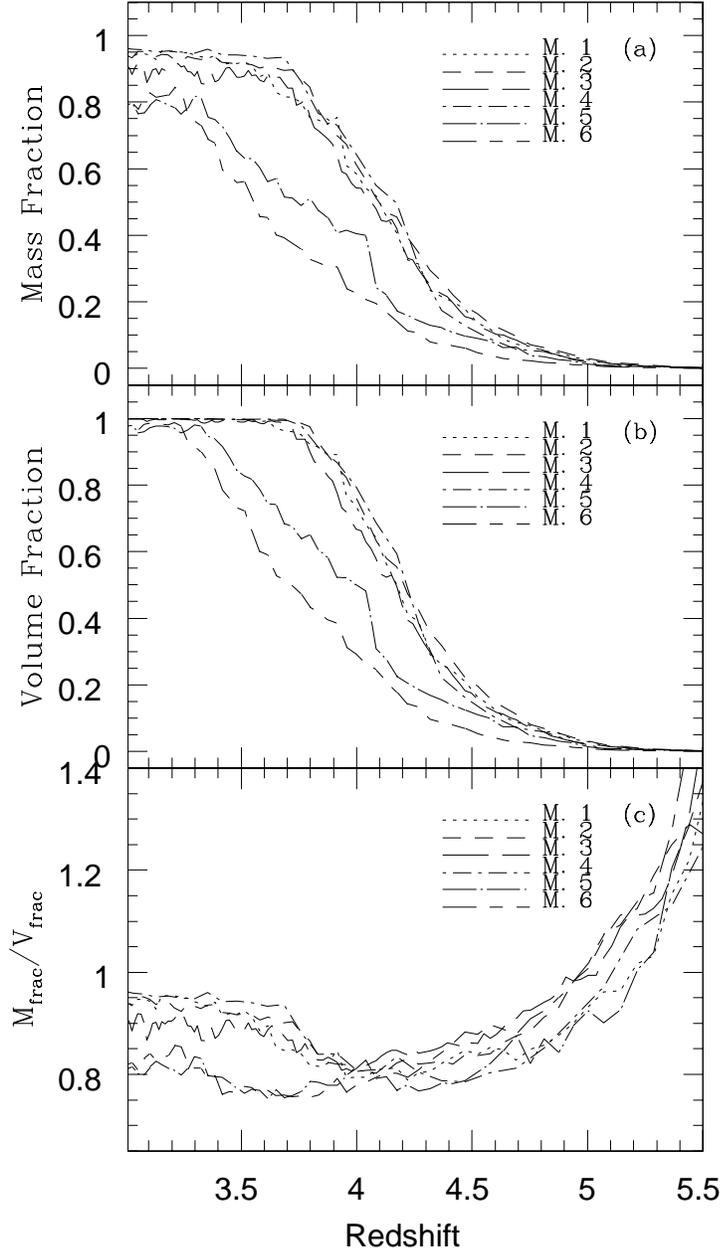}}
\end{picture}
\caption{Evolution of the ionized mass fraction (a), ionized volume
fraction (b), and the ionized mass to volume ratio (c) as functions of
redshift for each model.  In each case, full volume ionization of 
the Helium {\small II} is essentially attained by
a redshift $z\simeq 3.3$ with the volume fraction being
consistently larger than the mass fraction for $z\lesssim 5$ (see
text for discussion).}
\end{figure}

\clearpage

\begin{figure}[htb]
\figurenum{4}
\setlength{\unitlength}{1in}
\begin{picture}(6,6.5)
\put(-0.65,-1.7){\includegraphics{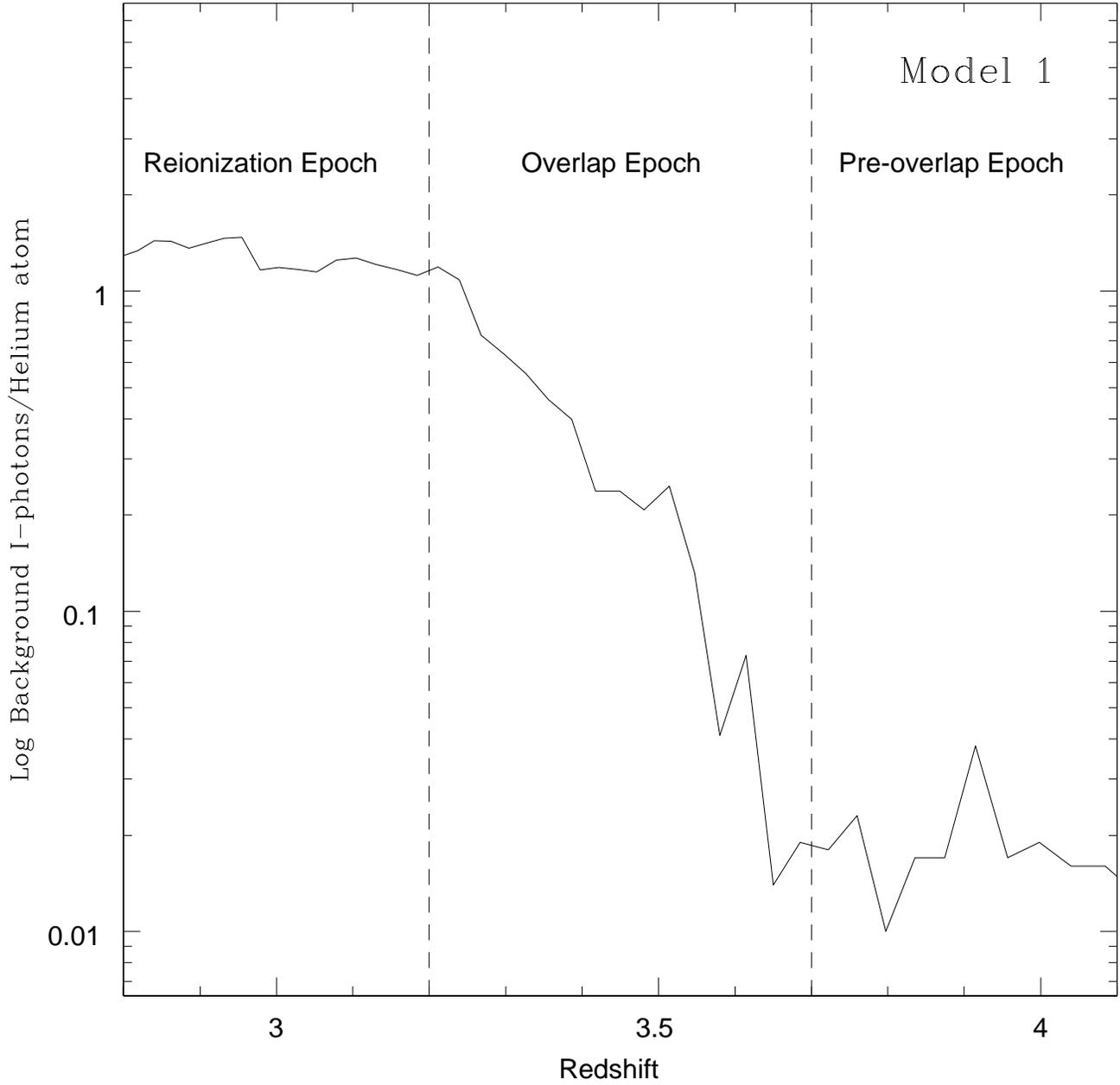}}
\end{picture}
\caption{Evolution in the number of ionizing photons per helium atom
available in the diffuse background component of the radiation
field. The region between the vertical dashed lines represents a short
fraction of a Hubble time where the intensity dramatically
increases, characteristic of an overlap epoch (e.g. Gnedin 2000).}
\end{figure}

\clearpage

\begin{figure}[htb]
\figurenum{5}
\setlength{\unitlength}{1in}
\begin{picture}(6,6.5)
\put(-0.65,-1.7){\includegraphics{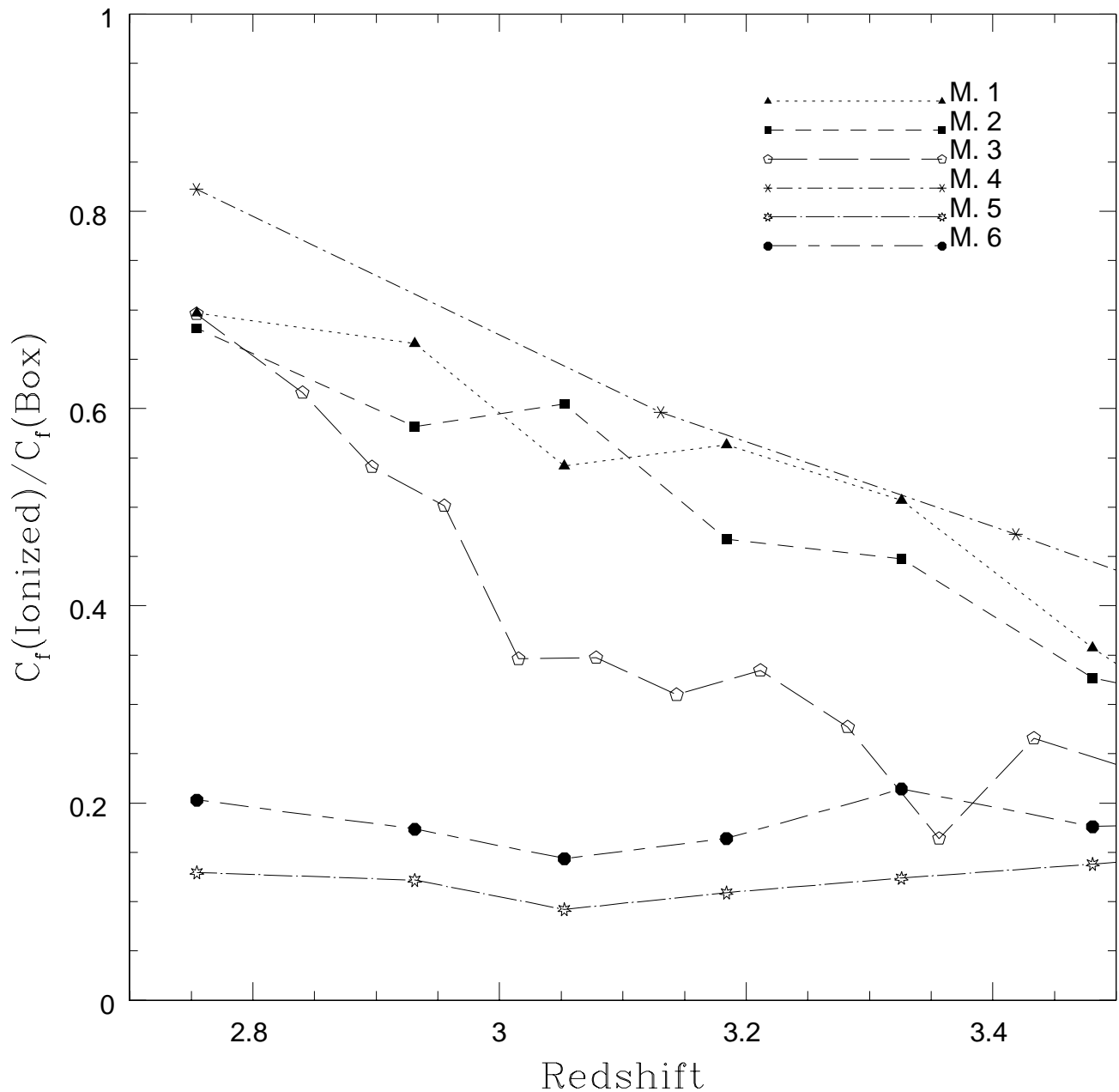}}
\end{picture}
\caption{Evolution of the ratio of the clumping factor within regions
that have been highly ($>90\%$) ionized in mass to the ratio of the
clumping factor of the entire volume. The low levels for model 5
and 6 reflect their inability to keep the densest cells ionized.  The
general uptrend towards lower redshifts seen in the other models is
consistent with ionization zones expanding into the densest structure
harboring relatively more clumped gas.}

\end{figure}

\clearpage

\begin{figure}[htb]
\figurenum{6}
\setlength{\unitlength}{1in}
\begin{picture}(6,6.5)
\put(-0.65,-3.1){\includegraphics{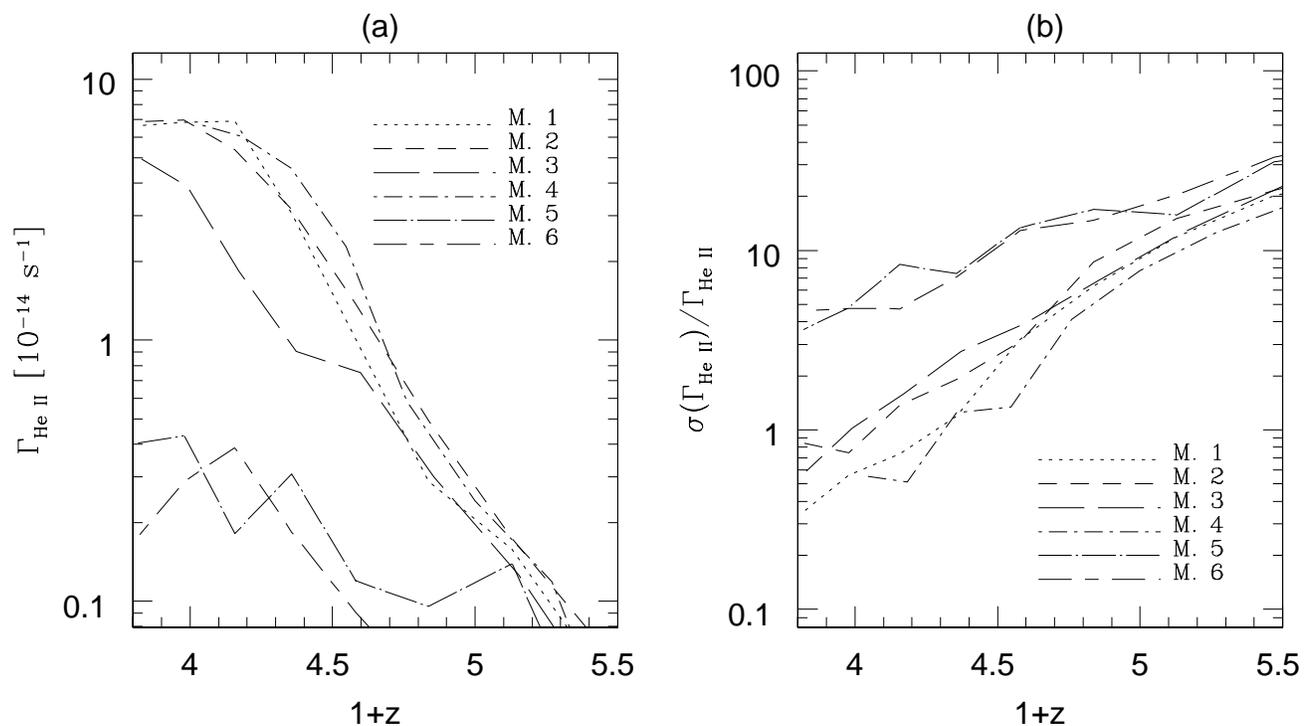}}
\end{picture}
\caption{Evolution of the mean volume-weighted photoionization rate in
the simulation volume ({\it left}) and corresponding dispersion ratio
({\it right}) for each model. Mean rates were obtained by averaging
over all cells in the grid.}
\end{figure}

\clearpage
\begin{figure}[htb]
\figurenum{7}
\setlength{\unitlength}{1in}
\begin{picture}(6,6.5)
\put(-0.65,-1.9){\includegraphics{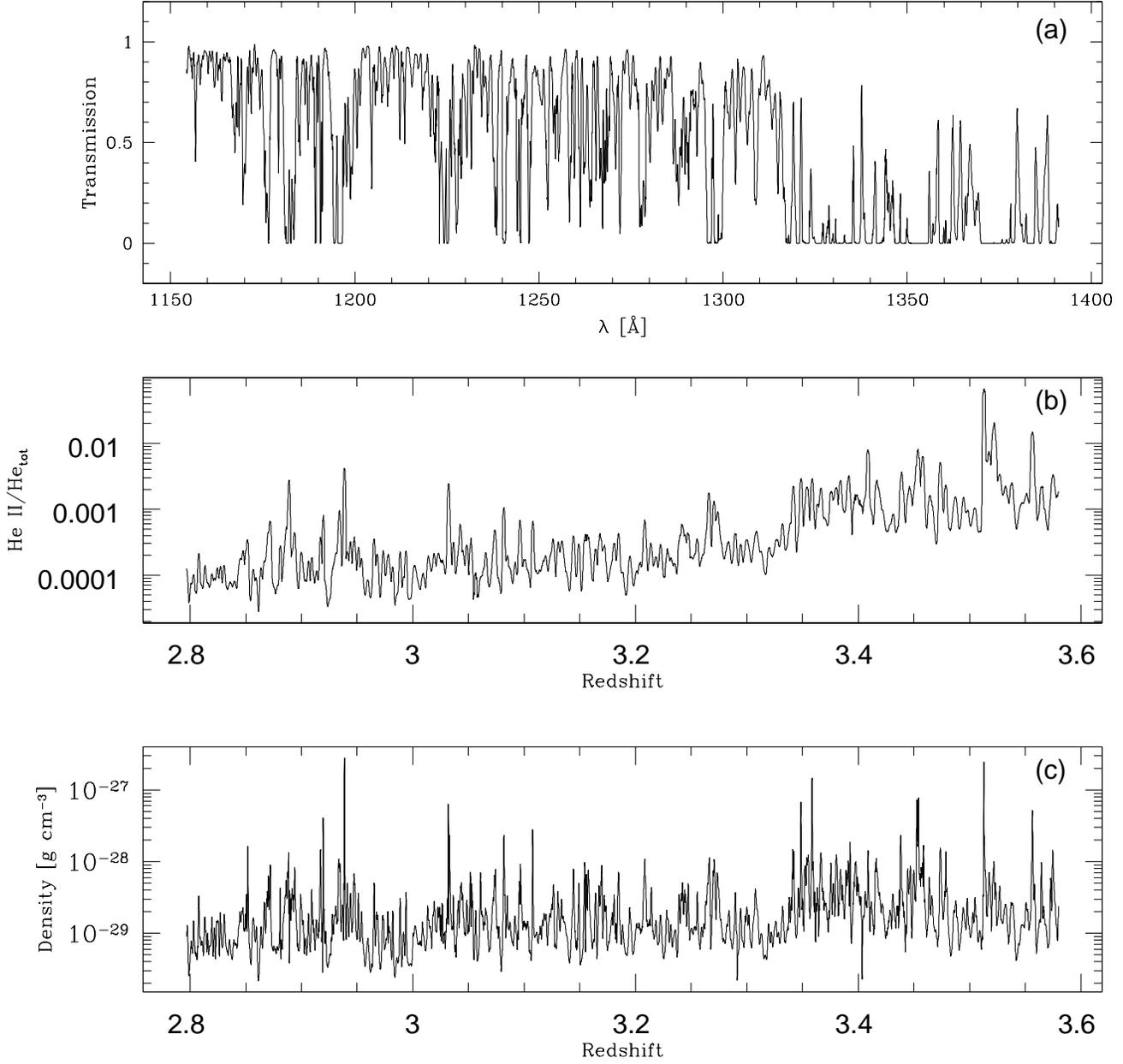}}
\end{picture}
\caption{Sample transmission spectrum (a) from model 1 spanning
$\lambda=$1150-1400 \AA \ along with the corresponding HeII
fraction (b) and gas density (c) along the LOS out to $z=$3.6.}
\end{figure}

\clearpage
\begin{figure}[htb]
\figurenum{8}
\setlength{\unitlength}{1in}
\begin{picture}(6,6.5)
\put(-0.65,-1.9){\includegraphics{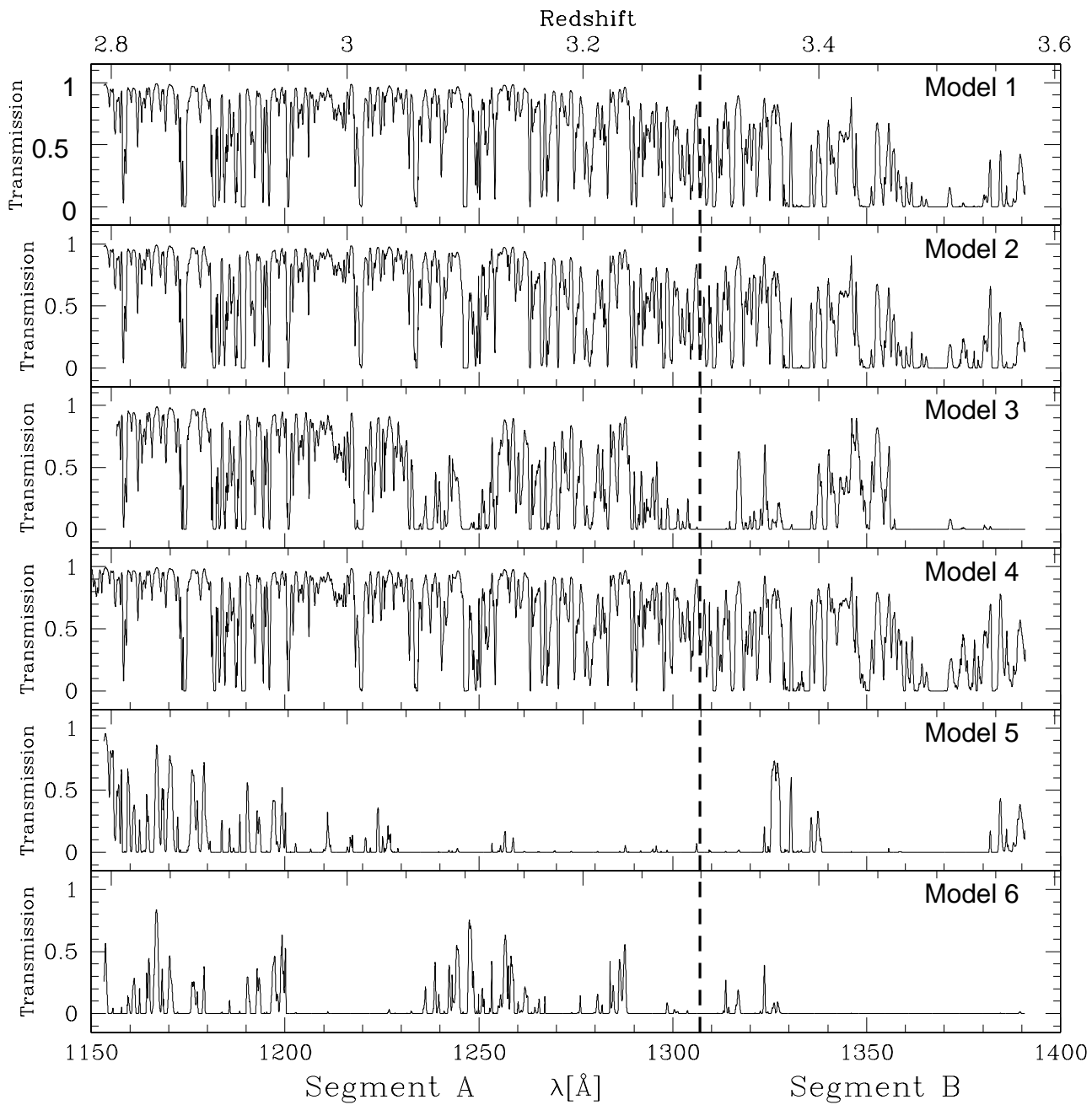}}
\end{picture}
\caption{Transmission spectra along a particular LOS out to $z=$3.6
extracted from the six models.  The dashed vertical line delineates 
where the spectra were divided for analysis purposes.}
\end{figure}

\clearpage

\clearpage
\begin{figure}[htb]
\figurenum{9}
\setlength{\unitlength}{1in}
\begin{picture}(6,6.5)
\put(-0.65,-3.7){\includegraphics{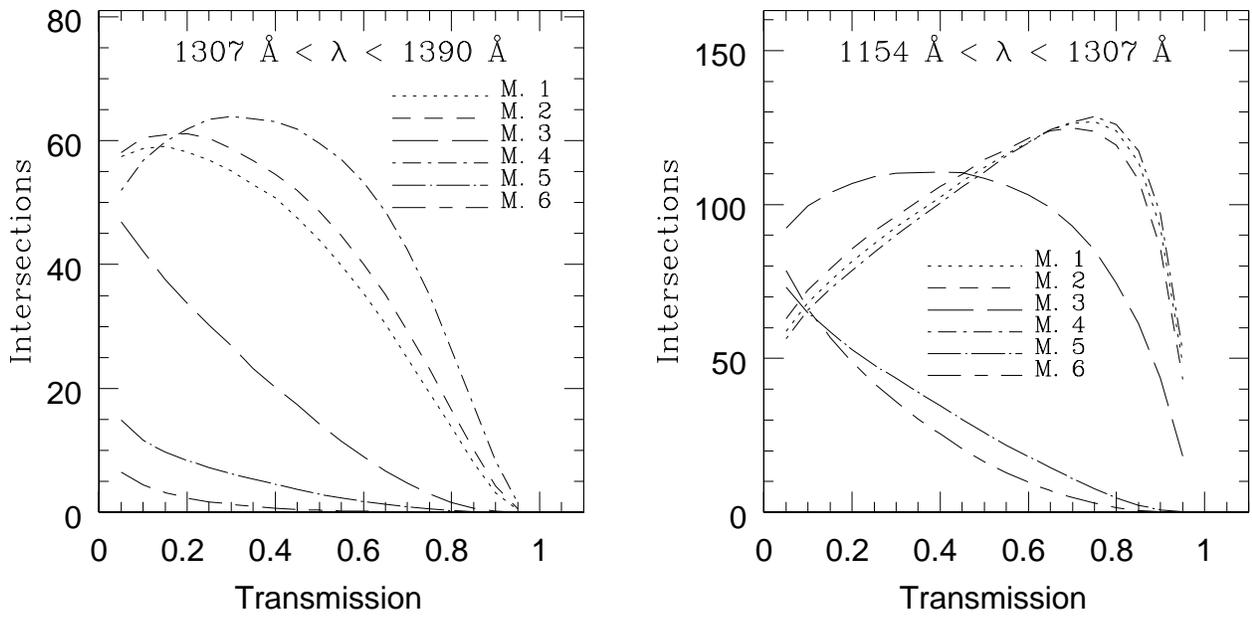}}
\end{picture}
\caption{Number of intersections through segment B ({\it left}) and segment
A ({\it right}) of the spectra of each model as a function of
transmission. Note the strong contrast between models within each
regime and the evolution of the curves between the regimes.}
\end{figure}

\clearpage
\begin{figure}[htb]
\figurenum{10}
\setlength{\unitlength}{1in}
\begin{picture}(6,6.5)
\put(-0.65,-3.7){\includegraphics{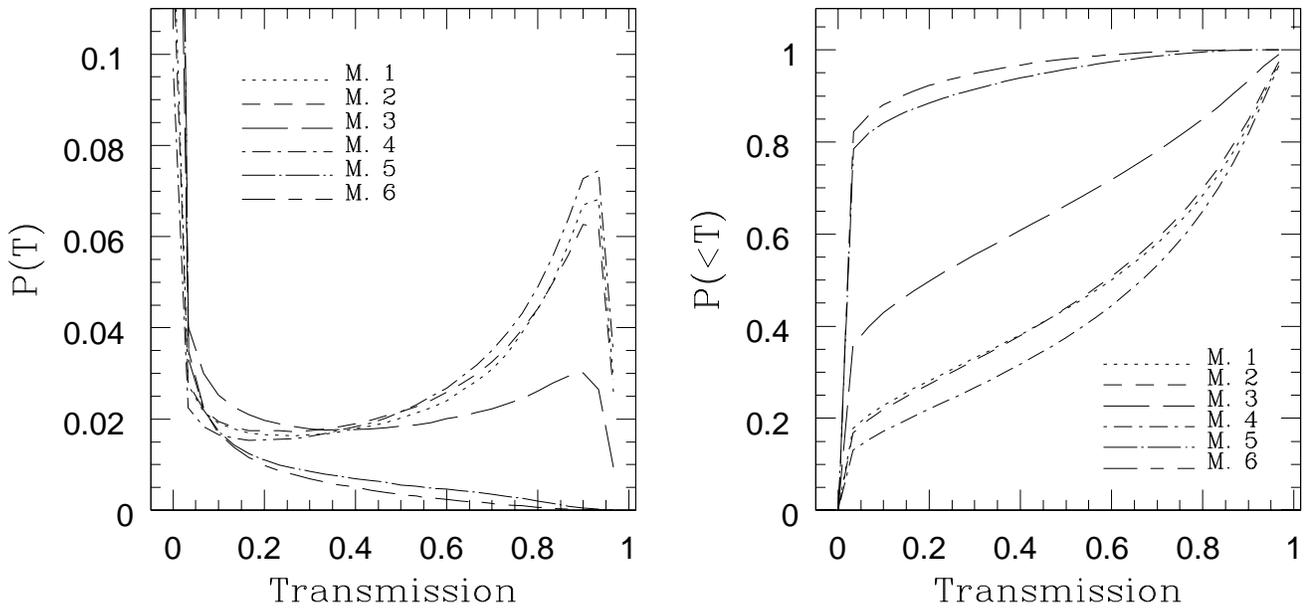}}
\end{picture}
\caption{TPDF ({\it left}) and the cumulative TPDF ({\it right}) for
each of the models averaged over 500 spectra.}
\end{figure}

\clearpage
\begin{figure}[htb]
\figurenum{11}
\setlength{\unitlength}{1in}
\begin{picture}(6,6.5)
\put(-0.65,-1.9){\includegraphics{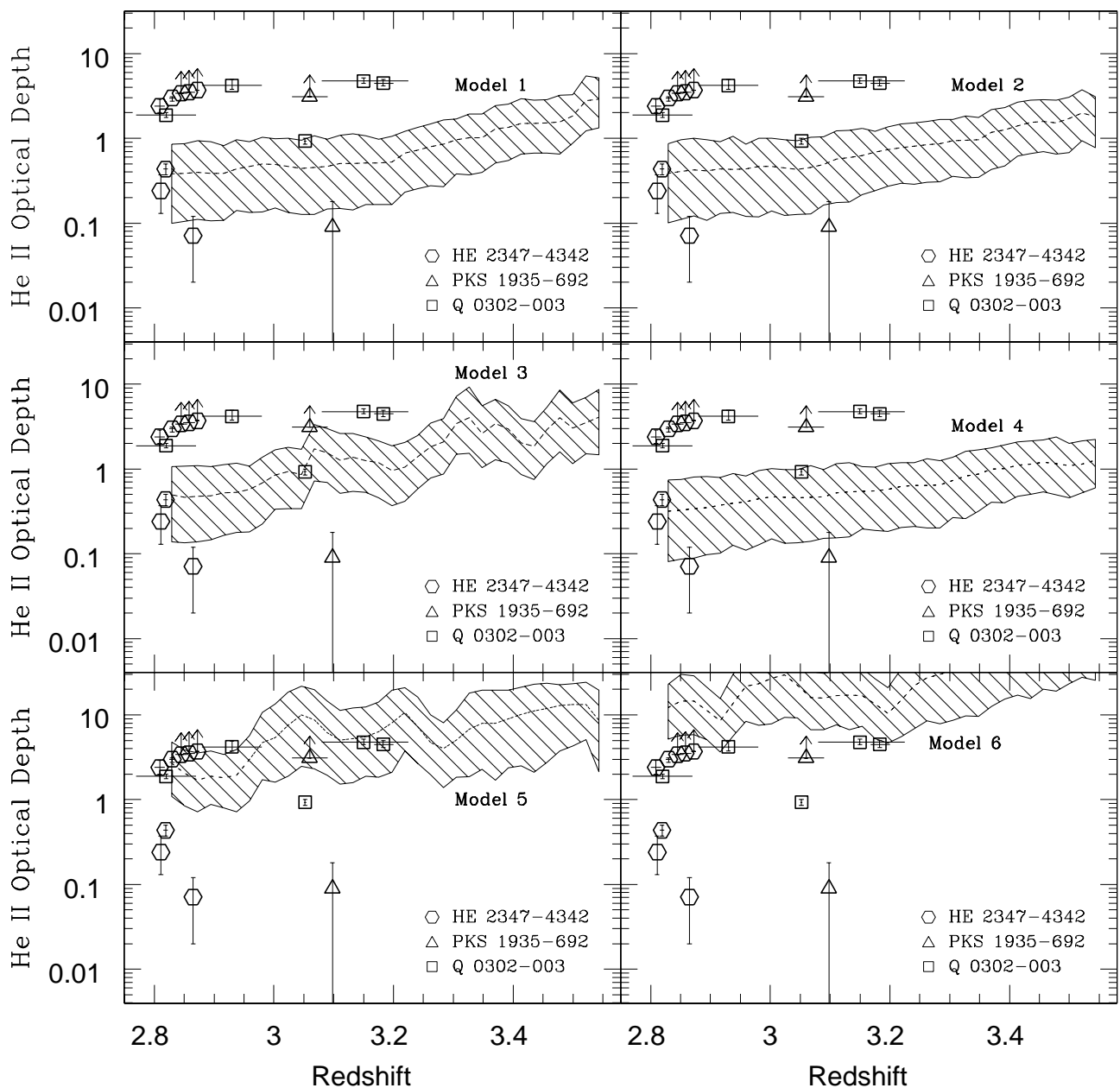}}
\end{picture}
\caption{Redshift evolution of the effective mean optical depth of He
{\small II} absorption. Hatched regions represent the optical depth
derived from the simulations at the 95\% confidence level with the
dashed lines indicating mean values. Observational results from
quasars HE 2347-4342, PKS 1935-692 and Q 0302-003 are plotted for
comparison.}
\end{figure}

\clearpage

\begin{figure}[htb]
\figurenum{12}
\setlength{\unitlength}{1in}
\begin{picture}(6,6.5)
\put(-0.65,-3.7){\includegraphics{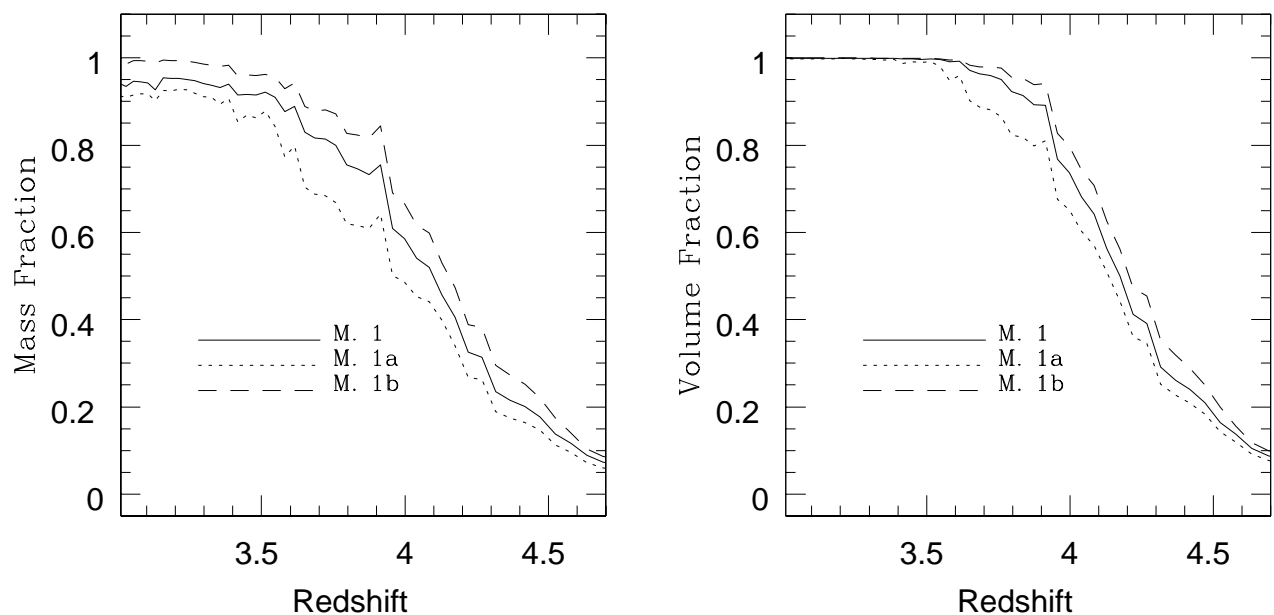}}
\end{picture}
\caption{Comparison of the ionized mass fraction ({\it left}) and
ionized volume fraction ({\it right}) as a function of redshift
between models 1, 1a, and 1b. Models 1a and 1b are identical to model 1
except for the ionized gas temperatures and clumping factors,
respectively. In model 1a the ionized gas temperature was set to $7000
\ K$ and in model 1b, clumping factors were all set to unity.}

\end{figure}

\clearpage

\begin{figure}[htb]
\figurenum{13}
\setlength{\unitlength}{1in}
\begin{picture}(6,6.5)
\put(-0.65,-1.7){\includegraphics{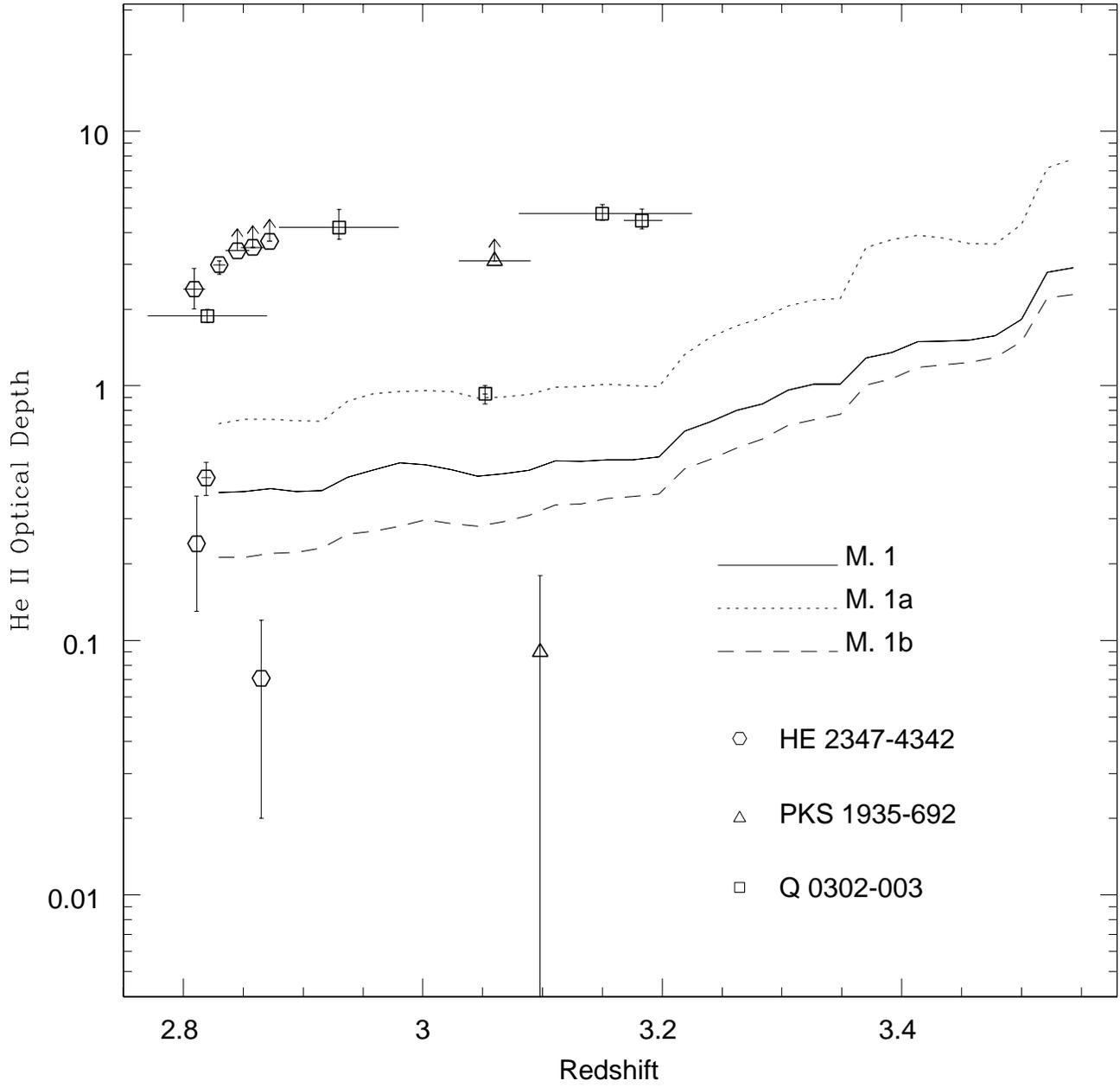}}
\end{picture}
\caption{Comparison of the redshift evolution of the effective mean
optical depth of He {\small II} absorption between models 1, 1a and
1b.}
\end{figure}

\clearpage

\end{document}